\documentclass{icrc2009}

\usepackage{graphicx}   
\usepackage{caption}    
\usepackage[font=footnotesize]{subfig} 
\usepackage{fixltx2e}
\usepackage{url}

\newcommand{\shorttitle}[1]%
{\markboth{Proceedings of the 31\MakeLowercase{$^{st}$} ICRC, {\L}\'{o}d\'{z} 2009}{#1} }

\begin{document}
\title{The Andyrchy-BUST experiment: primary spectrum and composition around the knee}

\author{\IEEEauthorblockN{V.B.\,Petkov\IEEEauthorrefmark{1}\\
              for the Andyrchy-BUST collaboration}
                            \\
\IEEEauthorblockA{\IEEEauthorrefmark{1}Institute for Nuclear Research of RAS, Russia }
 }

\shorttitle{V.B.\,Petkov The Andyrchy-BUST experiment: ...} \maketitle

\begin{abstract}
 The main goal of the Andyrchy-BUST experiment is to study the primary cosmic rays  spectrum and
composition around the knee. The experimental data on the knee, as observed in the electromagnetic and high
energy muon components, are presented. The electromagnetic component in our experiment is measured using
the "Andyrchy" EAS array. High energy muon component (with 230 GeV threshold energy of muons) is measured
using the Baksan Underground Scintillation Telescope (BUST). The location of the "Andyrchy" right above the
BUST gives us a possibility for simultaneous measurements of both EAS components.

  \end{abstract}

\begin{IEEEkeywords}
 EAS, primary spectrum, primary composition
\end{IEEEkeywords}

\section{Introduction}
In the range of primary energies of $10^{14} - 10^{15}$ eV per nucleus, direct methods for studying the
energy spectrum and nuclear composition of primary cosmic rays become inefficient because of a decrease in
the flux of primary particles with an increase in their energy. Therefore, at these and, of course, higher
energies, indirect methods based on simultaneous measurement of the characteristics of different components
of extensive air showers (EASs), which are initiated by the primary particle in the atmosphere, are used.
But the interpretation of these measurements requires their comparison with EAS simulations in the
atmosphere. In turn, the calculation results depend on the hadronic interaction models. The main problem is
the extrapolation of these models into kinematical and energy regions still unexplored by present-day
collider experiments. So, the measurements of different EAS components are now used for both studying the
primary composition and testing interaction models \cite{bib:KASCADE99} -  \cite{bib:KASCADE09}.


In this paper three types of experimental data are analyzed: muon number spectrum, EAS size spectrum and
correlation between muon number and EAS size simultaneously measured. Integral muon number spectrum has
been measured using the Baksan Underground Scintillation Telescope (BUST) \cite{bib:Petkov2008}.
 The  EAS size spectrum has been measured using the "Andyrchy" EAS array \cite{bib:Chudakov97}.
 The dependence of the mean number of high energy muons on EAS size has been measured
 by simultaneous operation of both devices  \cite{bib:Chudakov97a}, \cite{bib:Petkov2003}.\\

\section{Facilities}
\begin{figure}[!t]
  \centering
  \includegraphics[width=1.0 \columnwidth ]{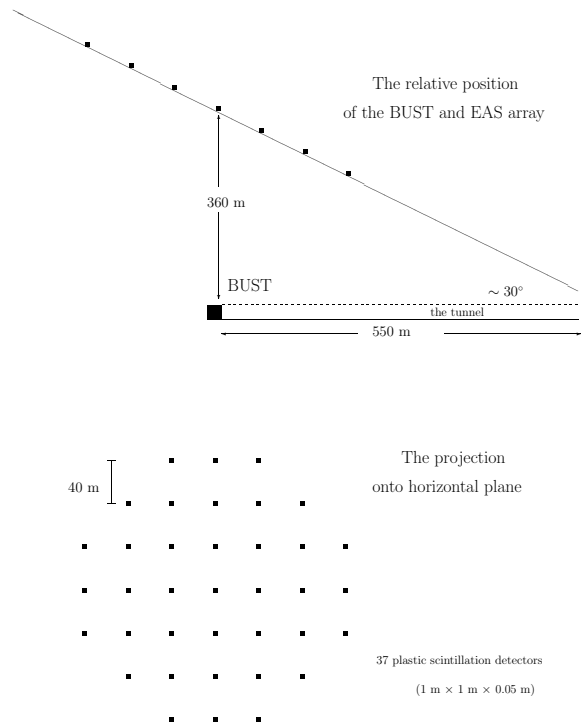}
  \caption{Andyrchy EAS array.}
  \label{Andyrchy}
 \end{figure}
 The "Andyrchy" EAS array is located on the slope of the Andyrchy mountain above BUST
(43.3$^{\circ}$ N, 42.7$^{\circ}$ E)  and consists of 37 plastic scintillation detectors. A plastic
scintillator of the 5 cm thickness has an area of $1 m^2$ and is viewed with a single PMT. The detectors
are designed for both timing measurements (for EAS arrival direction) and evaluation  of primary energy
(via EAS core localization and determination of total number of particles in shower). The distance between
the detectors is about 40 m in projection to the horizontal plane and the overall area of the installation
is $5\cdot10^4$ m$^2$; the solid angle at which the array is viewed from the BUST is 0.35 sr. The central
detectorof the array is above the BUST's center at a vertical distance of about 360 m and at 2060 m above
the sea level. The difference between the
heights of the upper and lower rows of detectors is $\sim$150 m.\\
The energy deposition measurement is performed in natural units, so called relativistic particles. One
relativistic particle (r.p.) is the most probable energy deposition from a single cosmic ray particle.
For our detector it is 10.6 Mev \cite{bib:Petkov2006}. The range of the energy deposition measurements
is from
0.5 r.p. (the threshold of the Charge-to-Time Converter) up to more than 1000 r.p.\\
Trigger formation and all measurements are performed in a registration
 room, which is placed near the center of the array (length of connection
 cables is up to 280 m).
The shower trigger condition requires signals from 4 detectors within 3 microseconds. The trigger's rate is
about 9 s$^{-1}$. The array and its characteristics are described in more details in \cite{bib:Petkov2006}.

BUST \cite{bib:Alekseev79} is a large device $16.7 \times 16.7\ m^2$ area and 11.1 m height), located in a
cave under mountain slope. The four vertical sides and four horizontal planes are completely covered with
standard liquid scintillation detectors. The standard detector consists of an aluminium tank with
$0.7\times 0.7\times 0.3\ m^3$ dimensions and is filled with liquid scintillator on the base of
white-spirit. Total number of the detectors is 3180. Every counter is viewed with one PMT. The construction
of  BUST allows one to reconstruct tracks of muons crossing the telescope. Coordinates of hit detectors is
used as input information for muon group parameters determination. The telescope allows one to determine
the number of passing muons, their coordinates (with 0.7 m accuracy) and the arrival direction (with $1.5$
degree accuracy). The coincidence trigger between "Andyrchy" and BUST is produced when one or more muons
crossing
 the telescope ($\approx 12s^{-1}$) coincide with the shower trigger within 51.2 microseconds; the
 coincidence rate is about $0.1s^{-1}$.

\begin{figure}[!t]
  \centering
  \includegraphics[width=1.0 \columnwidth ]{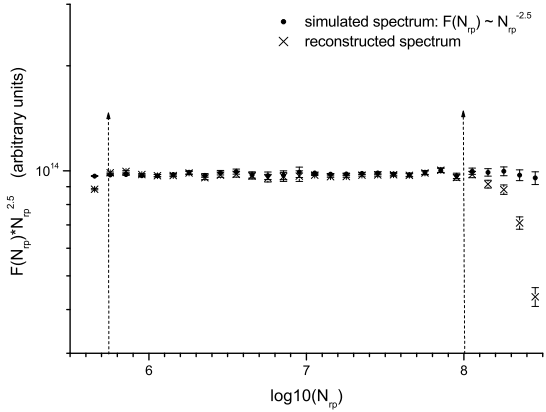}
  \caption{Simulated and reconstructed size spectra.} \label{Fig_2}
 \end{figure}

\section{EAS size spectrum}
The standard definition of the shower size $N_e$ is the total number of the charged particles (mainly
$e^{\pm}$) at the level of observations. As a scintillation detector measures the energy deposition, and
not the number of particles, the reconstruction of shower parameters is performed in units of relativistic
particles (r.p.). The measured size $N_{r.p.}$ is the total energy deposition in allegedly continuous
infinite detector. The shower size $N_{r.p.}$, the slope of the lateral distribution function and the core
location are determined by a $\chi^2$-like method, in which the logarithm of the energy deposition in each
detector is compared with the one expected from the NKG lateral distribution function
$$
 \rho(r)=N_{r.p.}{C(s)\over r_0^2} \left({r\over r_0}\right)^{(s-2)}
{\left(1+{r\over r_0}\right)}^{(s-4.5)}
$$
with $r_0 = 96$ m. The NKG function reproduces with a good accuracy the experimental
data~\cite{bib:Chudakov97}.

\begin{figure}[!t]
  \centering
  \includegraphics[width=1.0 \columnwidth ]{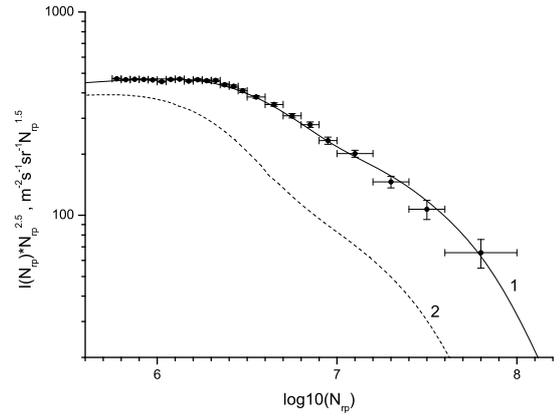}
  \caption{EAS size spectrum. Points - experiment. Lines 1 and 2 are calculated spectra for the primary
compositions 1 and 2, correspondingly.} \label{Fig_3}
 \end{figure}

In the present analysis, only the showers with:\\
1) $\sec \theta \le 1.05$ (near vertical events);\\
2) $\ge 22$ fired detectors;\\
3) $\ge 4$ detectors with energy deposition $E_d \ge 10$ r.p. well inside the array;\\
4) reconstructed axes in central part of the array (the distance from the center is not larger than 50 m)\\
were taken into account.\\
The accuracy of reconstruction was calculated using data obtained from a simulation that includes the
experimental dispersion. Figure \ref{Fig_2} shows the reconstructed size spectrum together with the
simulated one, the size spectrum is reconstructed without distortions for showers with $5.75 \le \lg
N_{r.p.} \le 8.0$. For these showers the accuracy of the $N_{r.p.}$ determination is better than 15\% and
the accuracy of the axis position determination is better than 5 m.
 Figure \ref{Fig_3} shows the measured differential size spectrum taken during live
time $1.297\cdot 10^8$ s (1501.2 days), in r.p. units. The steepening of the spectrum is observed at
$\lg N_{r.p.}\approx 6.35$.

\begin{figure}[!t]
  \centering
  \includegraphics[width=1.0 \columnwidth ]{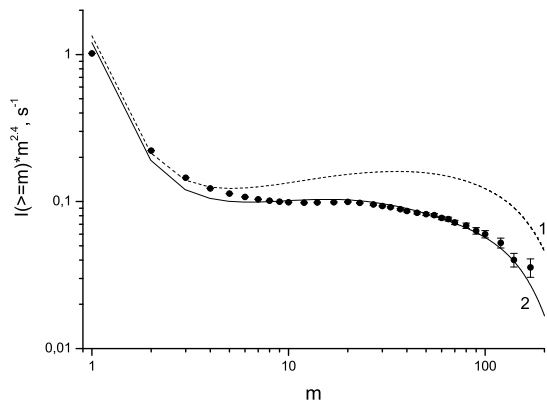}
  \caption{Integral muon tracks number spectrum. Points - experiment. Lines 1 and 2 are calculated
   spectra for the primary compositions 1 and 2, correspondingly.} \label{Fig_4}
 \end{figure}

\section{Muon number spectrum}
Coordinates of hit BUST detectors are used to reconstruct tracks of muons crossing the telescope. Generally
the number of muon tracks $m$ differs from the number of muons $m_\mu$ in the group. In the case where the
distance between muons is small enough (compared to the individual detector size) the number of
reconstructed muon tracks is smaller than the number of muons in the group. Opposite effect is also present
since interacting muons can produce particles which might increase the number of hit detectors, therefore
the number of reconstructed muon tracks in such a case can be larger than the number of muons in the group.
Furthermore there is some arbitrariness in muon track determination: for example, a track may cross two,
three or four telescope planes and so on. Hence it is necessary to convert the number of reconstructed muon
tracks to the number of muons in the group taking into account all mentioned effects. Conversion factors
depend on muon lateral distribution function. Lateral distribution might depends on muon energy
distribution in EAS and on primary nucleus energy (per nucleon) etc. In order to avoid additional
uncertainty we use only experimental muon tracks number spectrum for study of primary composition. The
conversion  of the number of muons  to the number of reconstructed muon tracks is included in the
calculations. The integral spectrum on the number of muon tracks for near vertical directions ($\theta \le
20^{\circ}$, effective muon threshold energy is 230 GeV) was measured for m = 1 - 250  ((Fig. \ref{Fig_4}).
This spectrum was obtained using two BUST data sets. The first data set has been taken during 2001 - 2004
(live time 3.3 years) and contains information about all the BUST events. The second data set has been
taken during 1984 - 1995 (live time 9.8 years) and contains information about BUST events where 100 and
more detectors were fired. The later condition corresponds to 20 or more muons crossing BUST for near
vertical directions. It is to be noted that owing to muon number tracks saturation there is no sense to
analyze this spectrum for $m > 170$, where the relation $m_\mu/m$ becomes more than 1.5.

\begin{table*}[t]
  \caption{Primary composition 1.}
  \centering

\begin{tabular}{|c|c|c|c|c|c|}
\hline
$Z$            & 1 & 2 & 6-8 & 10-16 & 20-26\\

$\overline Z$  & 1       & 2         & 7.2        & 12.7         & 25.2 \\

$\overline A$  & 1       & 4         & 14.4       & 25.5         & 54.2 \\

${F_z}^0, (m^2\ s\ sr\ TeV)^{-1}$ & 0.12  & 0.056  & 0.03  & 0.035  & 0.0267  \\

$\gamma_z$     & 2.75 & 2.64 & 2.66 & 2.70  & 2.62 \\
\hline
\end{tabular}
\end{table*}

\section{The mean number of muons vs. EAS size}
\begin{figure}[!t]
  \centering
  \includegraphics[width=1.0 \columnwidth ]{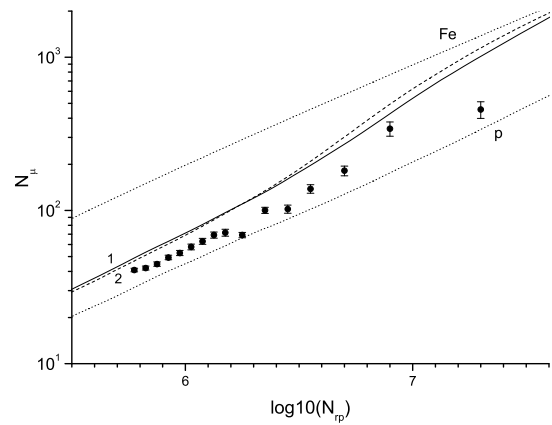}
  \caption{$\overline N_{\mu}(N_{rp})$ dependence. Points - experiment.
  Solid line 1 and dashed line 2 are calculated dependencies for the primary
compositions 1 and 2, correspondingly. The dotted lines are calculated dependencies for pure protons and
iron nuclei.} \label{Fig_5}
 \end{figure}

The size, axis position and the EAS arrival direction are determined using the "Andyrchy" array data;
the BUST data are used to determine the number of muons crossing BUST. The underground telescope
measures only a part of the total number of muons in EAS and the uncertainty in the determination of
the EAS axis position at the BUST level is comparable with the size of the BUST.  The mean number of
muons in BUST is determined as follows. Events within a given range of $N_e$ are grouped according to
the distance to the BUST's center with step $\delta R = 10$ m. For each group, the number of muons is

\begin{equation}
M(R_i) = \sum_{j=1}^{K_i}{m_{ij}}
\end{equation}

\noindent where  $K_i = K(R_i)$ is the number of EAS in the $i$-th group,
 and $m_{ij}$ is the total number of muons in the BUST in the $i$-th group for the $j$-th EAS.

Thus, the mean number of muons for a given range of $N_{r.p.}$ can be written as

\begin{equation}
{\overline n}(R_i) = \frac{M(R_i)}{K_i} = \frac{1}{K_i}\sum_{j=1}^{K_i}{m_{ij}}.
\end{equation}

 \noindent The mean number of muons in a shower is then calculated  as

\begin{equation}
{\overline N}_{\mu} = \frac{1}{S_t}\cdot \sum_i {\overline n}(R_i)\cdot S_r(R_i),
\end{equation}

\noindent where $S_t = 200$ m$^2$ is the effective area of the telescope, $S_r(R_i)$ is the area of the
ring of radius $R_i$ with $\delta R = 10$ m.

The fraction of muons in the telescope $\Delta(R)$ was measured as a function of the distance $R$ between
the center of the telescope and the EAS axis for a set of showers with a given value of $N_{r.p.}$. This
fraction for a given range of $N_{r.p.}$ is defined by

\begin{equation}
\Delta(R) = \frac{\overline n(R)}{\overline N_{\mu}(N_{r.p.})},
\end{equation}

\noindent where $\overline n(R)$ is the mean number of muons in the telescope at the distance $R$,
$\overline N_{\mu}(N_{r.p.})$ is the mean number of muons in EAS. Because $\Delta(R)$ depends on the
lateral distribution function (LDF) of high energy muons and BUST geometry only, LDF can be recovered from
these measurements. Preliminary results on high energy muons' LDF can be found in \cite{bib:Petkov2009}.

\begin{table*}[t]
  \caption{Primary composition 2.}
  \centering

\begin{tabular}{|c|c|c|c|c|c|}
\hline
$Z$            & 1 & 2 & 6-8 & 10-16 & 20-26\\

$\overline Z$  & 1       & 2         & 7.2        & 12.7         & 25.2 \\

$\overline A$  & 1       & 4         & 14.4       & 25.5         & 54.2 \\

${F_z}^0, (m^2\ s\ sr\ TeV)^{-1}$ & 0.102  & 0.079  & 0.018  & 0.018  & 0.02  \\

$\gamma_z$     & 2.77 & 2.65 & 2.70 & 2.70  & 2.64 \\
\hline
\end{tabular}
\end{table*}

\section{Calculations}

 The development of EAS in the Earth's atmosphere have been simulated by means of the CORSIKA code
(version 6900) \cite{bib:Heck98}. The QGSJetII-03 and Fluka were used as the high and low energy hadronic
interaction models. The CORSIKA output files were used then as input files for AndyrDet code, which
performs a detector response simulation. The results of the simulations were summarized as parametrization
functions of the EAS characteristics, that then was used for calculations of the observables (integral muon
number spectrum, EAS size spectrum, $\overline N_{\mu}(N_{rp})$ dependence). Vertically arriving CR
particles were used as primary CR particles. Simulations were performed for nuclei with atomic number $A =
1, 4, 14, 28, 56$. Primary energy per particle $E_0$ was taken from the range 10$^4$ GeV - 10$^{7.5}$ GeV
with a step of 0.5 of energy decade.

The integral muon number spectrum in BUST was calculated numerically in the same way as in
\cite{bib:Petkov2008}. This calculation method needs only such characteristics of high energy muon
component of EAS as: 1) $\overline N_{\mu}(E_0, A)$ -  muon production function (MPF) or mean number of
muons per EAS produced by nucleus with atomic number A and primary energy $E_0$; 2) $f(r, E_0, A) $ -
lateral distribution function (LDF); 3) $G(A, \overline N_{\mu}, N_{\mu})$ -  fluctuation function (FF).

The EAS size spectrum can be presented as:
\begin{equation}
I(N_{r.p.}) = \sum_{A} \int\limits_0^{\infty} \frac{dF_A(E_0)}{dE_0} W_A(E_0,N_{r.p.}) dE_0
\end{equation}
where ${dF_A(E_0)}/{dE_0}$ is the energy spectrum of the primary nuclei and $W_A(E_0,N_{r.p.})$ is the
probability for primary nucleus with energy $E_0$ to produce EAS with size $N_{r.p.}$ at observation level.

\begin{figure}[!t]
  \centering
  \includegraphics[width=1.0 \columnwidth ]{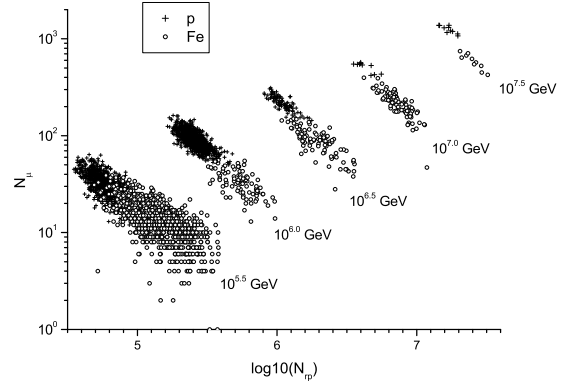}
  \caption{The number $N_{\mu}$ of high energy muons (with $E_{mu} \ge 230 GeV$) as function of $N_{rp}$
for the different primary energies of the primary protons and iron nuclei.} \label{Fig_6}
 \end{figure}

The dependence of the mean number of high energy  muons on EAS size was calculated taken into account the
energy spectra of the primaries and anticorrelation between the number of high energy muons in EAS and EAS
size at fixed primary energy (Fig. \ref{Fig_6}).

\begin{figure}[!t]
  \centering
  \includegraphics[width=1.0 \columnwidth ]{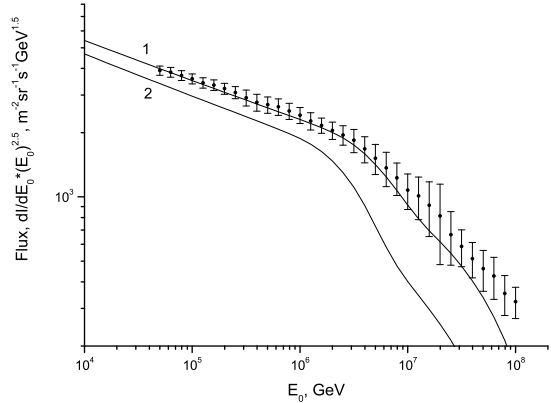}
  \caption{All-particle energy spectra: lines 1 and 2 - for compositions described in Tables 1 and 2,
correspondingly. Points - averaged all-particle energy spectrum from air shower experiments
\cite{bib:Horandel03}.} \label{Fig_7}
 \end{figure}

The calculations of the observables were performed for two composition models with five groups of primary
nuclei. Energy spectrum of every primary group is a power law with rigidity dependent knee
$E_{kz}=E_{kp}\cdot Z$:
\begin{equation}
\frac{dF_z(E_0)}{dE_0}={F_z}^0\cdot
{E_0}^{-\gamma_z}\left[1+\left(\frac{E_0}{E_{kz}}\right)^{\epsilon_c}\right]^{\frac{\gamma_z -
\gamma_c}{\epsilon_c}}
\end{equation}
where $E_0$ is energy per particle and ${F_z}^0$ is absolute flux at 1 TeV per particle
\cite{bib:TerAntonyan}, \cite{bib:Horandel03}. The sharp knee was applied for both composition models:
$\epsilon_c = 3.5$ and $\gamma_c = 5.2$. Both composition models do not give closest fit for the complete
data set of direct measurements, but the data do not contradict these models. All-particle energy spectra
for these composition are shown in Fig. \ref{Fig_7} together with averaged spectrum from air shower
experiments \cite{bib:Horandel03}.

The first composition is presented in Table 1, for this composition total flux  at 1 TeV is $0.2677\
(m^2\ s\ sr\ TeV)^{-1}$ and protons knee position is $E_{kp} = 4\cdot 10^3$ TeV. This composition
gives a good fit for the EAS size spectrum (Fig. \ref{Fig_3}, line 1). But the muon number spectrum
calculated for this composition is in contradiction with experiment (Fig. \ref{Fig_4}, line 1). For
the second composition (Table 2) the protons knee position is $E_{kp} = 2\cdot 10^3$ TeV and total
flux at 1 TeV is $0.215\ (m^2\ s\ sr\ TeV)^{-1}$. This composition gives a satisfactory fit for the
muon number spectrum (Fig. \ref{Fig_4}, line 2). But the EAS size spectrum calculated for this
composition is in contradiction with experiment (Fig. \ref{Fig_3}, line 2). For both compositions the
calculated $\overline N_{\mu}(N_{rp})$ dependences are very close to one another and both are in
contradiction with experiment (Fig. \ref{Fig_5}).

So, predictions from both considered mass composiiton models do not give satisfactory fits to all data
set obtained from our measurements (see Fig. \ref{Fig_3} and \ref{Fig_4}).

\section{High energy muon production function}
\begin{figure}[!t]
  \centering
  \includegraphics[width=1.0 \columnwidth ]{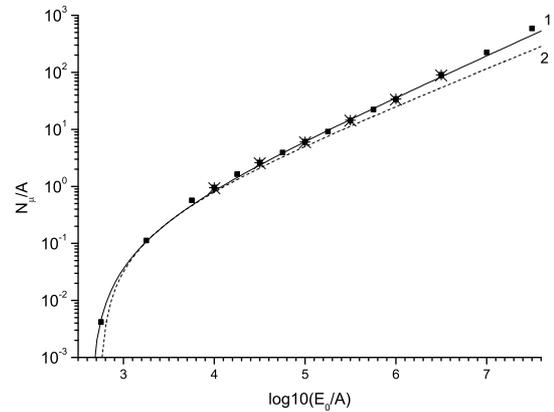}
  \caption{Muon production function. Points - CORSIKA results. Line 1 - fit of the CORSIKA results.
Line 2 - changed MPF} \label{Fig_8}
 \end{figure}

Discrepancies between EAS size spectrum and muon data can be lessened to a considerable degree if we
use another muon production function for muons with $E_{\mu} \ge 230$ GeV . The MPF can be expressed
by next formula:
\begin{equation}
\frac{\overline N_{\mu}}{A} = b \left[\left(\frac{E_0}{A}\right)^{\alpha} - c\right]^{\beta}
\end{equation}
Fit of the CORSIKA (QGSJetII-03) results was obtained for parameters: $b=0.0018$, $c=13.5$, $\alpha=0.43$
and $\beta=1.675$ (line 1 in Fig. \ref{Fig_8}).  It gives asymptotic behavior for $\overline N_{\mu} \sim
E_0^{0.720}$.

\begin{figure}[!t]
  \centering
  \includegraphics[width=1.0 \columnwidth ]{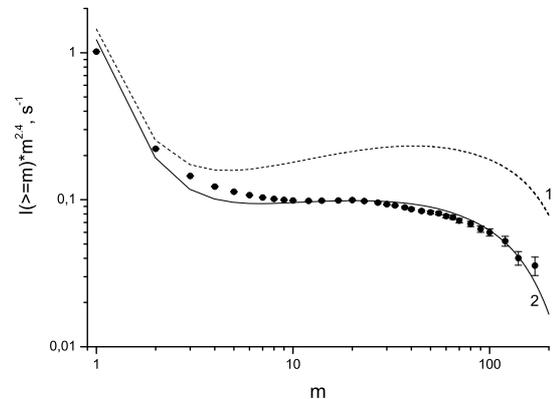}
  \caption{Integral muon tracks number spectrum. Points - experiment. Lines - calculated
   spectrum for the first primary compositions with: 1 - MPF from CORSIKA QGSJetII-03; 2 - changed MPF.}
\label{Fig_9}
 \end{figure}

The agreement between EAS size spectrum and muon data can be obtained for MPF with parameters: $b=0.0035$,
$c=14$, $\alpha=0.42$ and $\beta=1.54$ (line 2 in Fig. \ref{Fig_8}).  The asymptotic behavior of the mean
number of muons for this MPF is: $\overline N_{\mu} \sim E_0^{0.647}$. The integral muon number spectrum
calculated for the first primary composition and both MPF's is shown in Fig. \ref{Fig_9}.

\begin{figure}[!t]
  \centering
  \includegraphics[width=1.0 \columnwidth ]{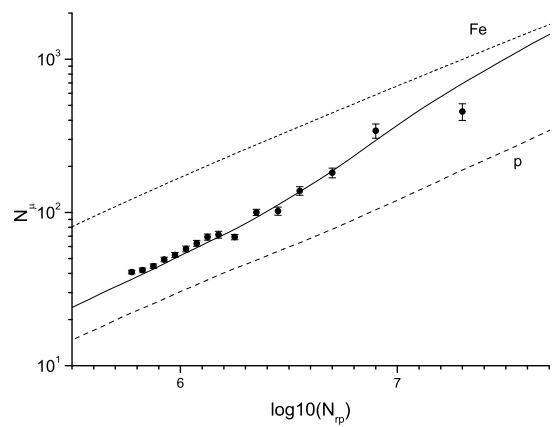}
  \caption{$\overline N_{\mu}(N_{rp})$ dependence. Points - experiment.
  Solid line is calculated dependencies for the first primary
compositions and changed MPF. The dashed lines are calculated dependencies for pure protons and iron nuclei
with changed MPF.} \label{Fig_10}
 \end{figure}

The $\overline N_{\mu}(N_{rp})$ dependences calculated using changed MPF are in Fig. \ref{Fig_10}. One
can see that the dependence calculated for the first primary composition and changed MPF gives
acceptable fit with experimental data.


\section{Conclusion}
It is widely known that none of the present interaction models can completely describe a full set of
experimental data for cosmic rays. Joint analysis of the various characteristics of different EAS
components, especially measured in one and the same experiment, can be used for both studying the
primary composition and testing interaction models.

In this paper three types of experimental data, taken in our experiment, were analyzed: high energy
($E_{\mu} \ge 230$ GeV) muon number spectrum, EAS size spectrum and dependence of the mean number of
high energy muons on EAS size. CORSIKA code v.6900, with QGSJetII-03 and Fluka as the high and low
energy hadronic interaction models, has been used for EAS simulations. As it was mentioned above, the
calculations of these observables need only the following characteristics of EAS:\\ 1)
$W_A(E_0,N_{r.p.},N_{\mu})$ - the probability for primary nucleus with atomic number A and energy
$E_0$
to produce EAS with size $N_{r.p.}$ and total number of high energy muons $N_{\mu}$ at the observation level;\\
2) $W_A(E_0,N_{r.p.})$ - the probability to produce EAS with size
$N_{r.p.}$ at the observation level;\\
3) $\overline N_{\mu}(E_0, A)$ -  high energy MPF;\\
4) $G(A, \overline N_{\mu}, N_{\mu})$ - high energy muons fluctuation function;\\ 5) $f(r, E_0, A)$ -
high energy muons LDF.\\
It should be noted that characteristics 2 - 4 can be derived from 1.

\noindent The analysis has shown that:\\ 1) our experimental data can be brought into good enough
agreement using changed MPF (Fig. \ref{Fig_8});\\ 2) in both cases the primary composition gets
heavier across the knee.

\vspace{3mm}

 {\bf Acknowledgements.}  The author is grateful to the Organising Committee of the 31$^{st}$
ICRC for the possibility to present this talk at the highlight session. This work was supported by the
"Neutrino Physics and Astrophysics" Program for Basic Research of the Presidium of the Russian Academy
of Sciences and by "State Program for Support of Leading Scientific Schools" (grant no.
NSh-321.2008.2). This work was also supported in part by the Russian Foundation for Basic Research,
Grant No. 08-07-90400.

\vspace{10mm}

\begin{center} \underline {The Andyrchy-BUST collaboration}

\vspace{2mm} V.B.\,Petkov$^1$, J.\,Szabelski$^2$, I.A.\,Alikhanov$^1$, A.N.\,Gaponenko$^1$,
Zh.Sh.\,Guliev$^1$, I.M.\,Dzaparova$^1$, V.I.\,Volchenko$^1$, G.V.\,Volchenko$^1$, A.F.\,Yanin$^1$

\vspace{1mm} $^1$Institute for Nuclear Research of RAS, Russia

\vspace{1mm}$^2$The Andrzej Soltan Institute for Nuclear Studies, Poland

 \end{center}

\end{document}